# Multirate Spectral Domain Optical Coherence Tomography

**Prabhav Gaur [1,*], Andrew Grieco [2] and Yeshaiahu Fainman [1]**

[1] Jacobs School of Engineering, University of California San Diego
[2] Qualcomm Institute, University of California San Diego
* Correspondence: pgaur@eng.ucsd.edu;

**Abstract:** Optical coherence tomography is state-of-the-art in non-invasive imaging of biological structures. Spectral Domain Optical Coherence Tomography is the popularly used variation of this technique, but its performance is limited by the bandwidth and resolution of the system. In this work, we theoretically formulate the use of phase modulators and delay lines to act as filters on the tomography system and scan multiple channels. Various channels are then combined in a digital computer using filter bank theory to improve the sampling rate . The combination of multiple channels allows for increasing the axial resolution and maximum unambiguous range beyond the Nyquist limit. We then simulate the multirate spectral domain optical coherence tomography with 2 channels. We show that a single delay line can improve the axial resolution while a pair of phase modulators can improve the maximum unambiguous range of the system. We also show the use of multirate filter banks to carry out this process. Thus, by using a few extra components in the spectral domain optical coherence tomography, its performance can be increased manifold depending on the number of channels used. The extra cost is the time taken to perform the extra scans that is trivial for stationary objects like biological tissues.

## 1. Introduction

Optical Coherence Tomography (OCT) is an important and powerful 3D imaging tool for several biomedical applications. The first OCT was demonstrated three decades ago by [1], but has evolved since then and has remained a major technique to study layered structures for ophthalmology. OCT was initially performed as Time Domain OCT (TD-OCT) [2] which is analogous to ultrasound. TD-OCT uses an echo time delay of backscattered light to create a cross-sectional image of the tissue under investigation. With the improvement in hardware technology, TD-OCT led to Fourier Domain OCT (FD-OCT)[3] which relies on multiple wavelengths to make the axial scan, also called the A-scan. The FD-OCT can be implemented in two different ways, namely Spectral Domain OCT (SD-OCT) [4] and Swept Source OCT (SS-OCT) [5]. Both of them use interferometry of reflected and reference light wave, and measurement of several wavelengths to gather depth information. SD-OCT works with a partially coherent broadband source and a spectrometer is used to detect the interference of different wavelengths. SS-OCT utilizes a wavelength sweep from a coherent laser and the interference is detected using a photodetector. The Fourier transform of the detected signal yields the depth information in both of the above cases. The A-scan can measure up to a depth of few mm with a resolution of few µm [6,7]. Multiple A-scan while moving in the other two dimensions, also called B-scan, can be used to determine the 3D structure of the object.

SD-OCT is currently one of the most popular commercial techniques [8] to evaluate tissue structure due to its high data acquisition rate, high axial resolution, good SNR and the simplicity of the hardware required to perform A-scan. SD-OCT has most extensively been used to diagnose diseases and disorders in the brain [9] and retina [10-12] but has also found application for other tissues such as the breast [13], kidney [14,15], and skin [16]. Recently, a wide variety of modifications have been applied to SD-OCT to improve its performance [17-30]. Given enough SNR [31] in the SD-OCT system, the axial resolution and the maximum unambiguous range are determined by the Nyquist-Shannon sampling theorem [32], where frequency and time (length) domain form a fourier pair. Let $f_o$ be the resolution of the spectrometer being used and $B$ be the bandwidth of the system either limited by source or spectrometer. The axial resolution $l_o$ and maximum unambiguous distance $L$ can be given by

$$l_o = c/B \quad ; \quad L = c/2f_o \qquad (1)$$

where $c$ is the speed of light in vacuum. Thus, the system after detection behaves like an analog system before and then like a digital system after photodetection.

In this work, we theoretically formulate the use of delay line and phase modulators along with multiple A-scan at the same position to improve the resolution and maximum unambiguous depth of SD-OCT. We show that the use of optical components at various positions in an SD-OCT system can be treated as filters acting on the depth information.

By making several scans with different filters, multiple channels can be created that have depth information encoded in them. After detection, the theory of multirate filter banks [33] is used to combine the channels and increase the resolution in the frequency or length domain. We then simulate simple cases of this system to demonstrate the multirate SD-OCT. This work is based on the principles of linear digital signal processing and statistical optics, which although well known, have not been combined in such a way to improve SD-OCT to the best of our knowledge.

## 2. Materials and Methods

The Multirate filter banks are sets of filters, decimators, and interpolators used widely in conventional digital systems [34]. Usually, decimators downsample the signal after passing through analysis filters. This compressed information is stored or transmitted via a channel. On the other end of the channel, the signal is interpolated or upsampled and passed through synthesis filters to retrieve the original information. The downsampling process means decreasing the system's resolution, which is similar to an undersampled tomography system. The tomography systems are also discrete after detection, and analog filters can be implemented by phase modulation/delay line on the optical carrier signal before detection and by digital processing after detection. Hence, the imaging system can be considered as a multirate filter bank with each scanning cycle representing a single channel and carrying object information in a compressed form. In this work, we formulate the underlying equations governing the SD-OCT to determine the analog filter applied to it. We use the theory of multirate filter bank to determine the digital filters needed to combine back the information. For proof of concept, we simulate a 2-channel filter bank implementation that results in a twofold improvement in both the length and frequency resolution of the tomography system. In our previous work [35], we have demonstrated theoretically and experimentally a similar multirate system for SS-OCT. For SS-OCT, a single phase modulator is sufficient to improve the frequency and length resolution. In this work we simulate a similar result for SD-OCT, where an extra phase modulator and delay line is needed. Also, the formulation requires the inclusion of the statistics of broadband source, which is not necessary for SS-OCT.

### 2.1. Spectral Domain OCT

SD-OCT uses a broadband source for illumination as shown in Figure 1(a). Let $r_s(t)$ be the complex field emitted by this source as a function of time $t$. We assume that $r_s(t)$ is an ergodic process, and thus by Wiener-Khinchin theorem [36] the power spectral density $S(f)$ of this source can be given by

$$PSD_s(f) = \mathbb{F}\{R_s(\tau)\}(f) = S(f) \tag{2}$$

where $\mathbb{F}\{.\}$ represents the Fourier transform, $f$ is the frequency and $R_s(\tau)$ is the autocorrelation function of the complex field emitted by the source given by

$$R_s(\tau) = \lim_{T \to \infty} \frac{1}{T} \int_{t=-T/2}^{T/2} r_s(t+\tau) r_s^*(t) \, dt \tag{3}$$

Next, consider a heterogeneous object with multiple optical media and their corresponding surface present only at spacings of effective optical distance $il_o$ from the first surface (Figure 1(a)). Since a reflected beam will pass through each section twice (once in the transmission direction, and once in the reflected direction), the effective optical path length of each section is defined as twice the distance multiplied by the effective index of the medium. $i$ is an integer in interval [1, $N$-1], where $N$ is the total number of surfaces that can be present, including the calibrated first surface which lies on the balance point of the interferometer and is a free parameter determined by the path difference between the sample and reference arm.. $l_o$ determines the axial resolution of the imaging system. The
light source is incident on this sample. The reflection from the $i$th surface is given by the following relation

$$r_i(t) = a(i) r_s(t - t_i) \tag{4}$$

$a(i)$ is a complex number describing reflection from the $i$th surface. The reflection coefficient can be calculated from Fresnel equations. Theoretically, $a(i)$ can have contributions from surfaces other than the $i$th surface. This is due to possible multiple reflections in between the surfaces in the multiple layers that give the same delay as the $i$th surface would have produced. But these extra terms can be neglected in biological samples with small refractive index changes because usually $r$ (reflection coefficient) $\ll t$ (transmission coefficient), which will attenuate the multiple reflections. This approximation was first used in Fizeau interferometer [37] and is often used in interferometry. If the $i$th surface is absent the $a(i)$ can be considered to be zero. $t_i$ is the time delay corresponding to reflection from the $i^{th}$ surface. Scattering is neglected to keep the formulation simple. The total reflection coming from the object is given by equation

$$r_{total}(t) = \sum_{i=1}^{N-1} r_i(t) \tag{5}$$

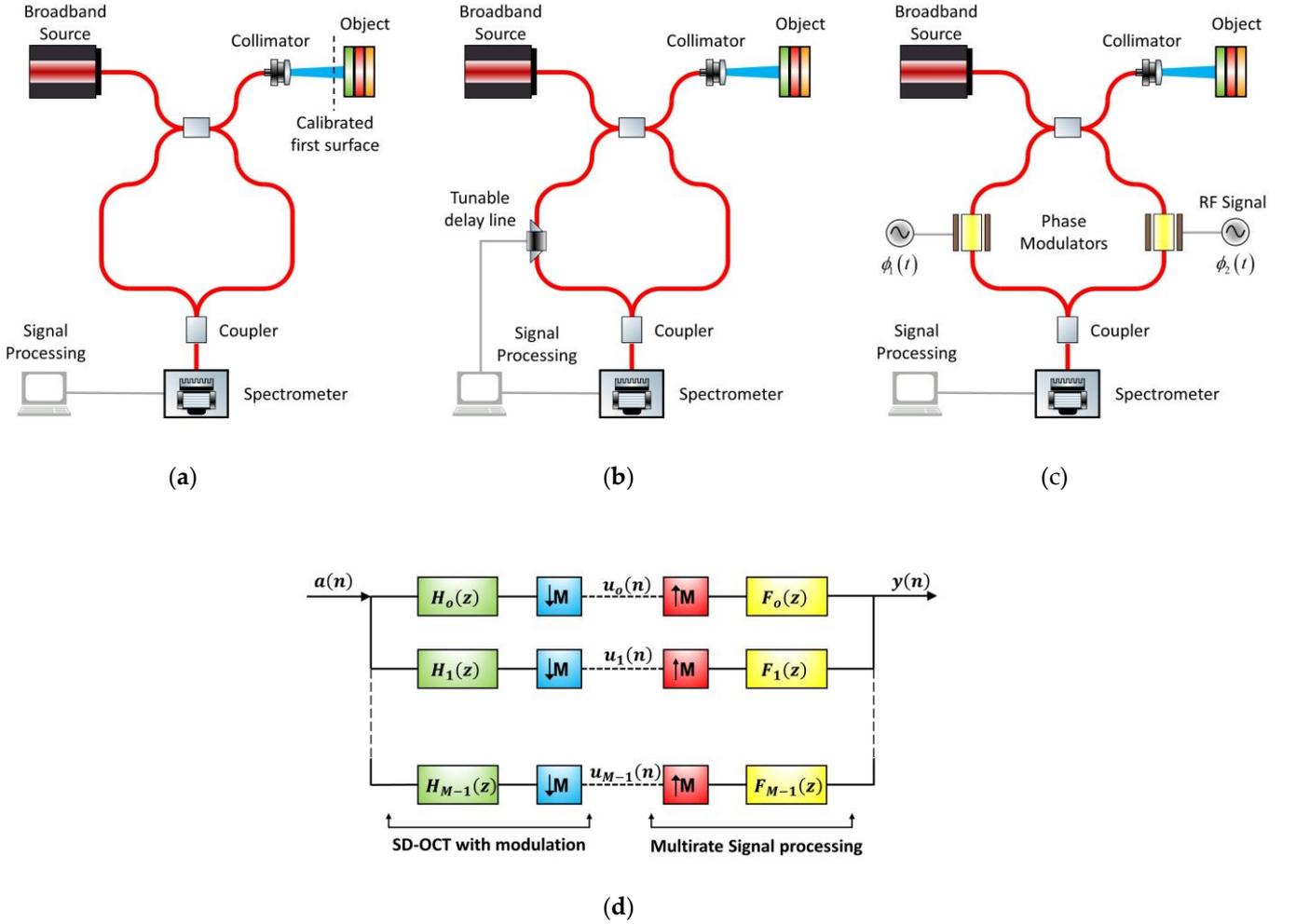

**Figure 1.** Setup for SD-OCT along with the variants studied in this work. (a) Regular SD-OCT setup in fiber and the object that is under investigation. (b) Addition of a delay line to SD-OCT in the sample arm. This implements a transfer function in the frequency domain and improves resolution in the length domain. (c) Addition of a couple of phase modulators to sample and reference arms. This implements a transfer function in the length domain and improves the maximum unambiguous range. (d) Representation of SD-OCT and the signal processing in form of a block diagram. The Z domain corresponds to the frequency or length domain depending on the implementation of (b) or (c). $a(n)$ represents the optical signal that the interference term of SD-OCT carries and our aim is to recover this signal digitally. The green blocks represent analysis filters that are provided optically using a delay line or phase modulators. The blue blocks represent downsampling which is naturally present in the system when the spectrometer has worse bandwidths or resolution than desired. Broken lines represent spectrum measurement using a spectrometer. The red blocks represent upsampling which is implemented on a digital computer. The yellow blocks represent synthesis filters that can be calculated from the theory of multirate signal processing and are also implemented digitally. Such scans are made M times with different filters as shown as M different channels. Finally, all the channels are combined to give $y(n)$ which should be close to a perfect reconstruction of $a(n)$ with desire resolution/maximum unambiguous range.

The field in the sample arm will be proportional to the $r_{total}$. The proportionality constant depends on 1) the coupling coefficient of the 3db fiber coupler, losses, etc. which are neglected as they correspond to scaling terms, and 2) sample arm length, which is assumed to be equal to that of the reference arm is also neglected.

$$r_{sample} = r_{total} = \sum_{i=1}^{N-1} r_i(t) \tag{6}$$

The field in the reference arm is the original field of the source that is transmitted to the object and is given by

$$r_{reference} = r_s(t) \tag{7}$$

The complex field at the spectrometer is

$$r(t) = r_s(t) + \sum_{i=1}^{N-1} a(i) r_s(t - t_i) \tag{8}$$

To calculate the power spectral density, we will use the Wiener-Khinchin theorem and assume that the statistics of the broadband source is ergodic. The autocorrelation function of this field is given by

$$R(\tau) = \lim_{T \to \infty} \frac{1}{T} \int_{t=-T/2}^{T/2} r(t+\tau) r^*(t) \, dt \tag{9}$$

$$R(\tau) = \lim_{T \to \infty} \frac{1}{T} \int_{t=-T/2}^{T/2} \left[ r_s(t+\tau) + \sum_{i=1}^{N-1} a(i) r_s(t - t_i + \tau) \right] \left[ r_s^*(t) + \sum_{i=1}^{N-1} a^*(i) r_s^*(t - t_i) \right] dt \tag{10}$$

The interference term is given by

$$R_{intf}(\tau) = \lim_{T \to \infty} \frac{1}{T} \int_{t=-T/2}^{T/2} \left[ \sum_{i=1}^{N-1} a(i) r_s(t - t_i + \tau) r_s^*(t) + \sum_{i=1}^{N-1} a^*(i) r_s(t+\tau) r_s^*(t - t_i) \right] dt \tag{11}$$

Thus, the interference term of the autocorrelation is given by

$$R_{intf}(\tau) = \sum_{i=1}^{N-1} a(i) R_s(\tau - t_i) + \sum_{i=1}^{N-1} a^*(i) R_s(\tau + t_i) \tag{12}$$

Assume a new set of time delay with mapping $t_i \to t_i$ for $i > 0$, $-t_{-i} \to t_i$ for $i < 0$, and $t_0 = 0$. Also mapping $a(i) \to a(i)$ for $i > 0$, $a^*(-i) \to a(i)$ for $i < 0$, and $a(0) = 0$ represents the calibrated first surface.

$$R_{intf}(\tau) = \sum_{i=-N+1}^{N-1} a(i) R_s(\tau - t_i) \tag{13}$$

The effective distance between the 1st and $i$th surface (as defined before) is $il_o$ for $i > 0$. Performing mapping for time delay $t_i$ as defined above, it can be shown that equation (14) holds for all possible values of $i$ in the interval $[-N+1, N-1]$

$$t_i = \frac{il_o}{c} \tag{14}$$

Applying Fourier transform on equation (13)

$$PSD_{intf}(f) = S(f) \sum_{i=-N+1}^{N-1} a(i) \exp\left( \frac{-j2\pi i f l_o}{c} \right) \tag{15}$$

where $j$ is the unit imaginary number. For a spectrometer like a grating spectrometer, we can measure discrete frequencies with resolution $f_o$. As we have $2N-1$ terms in the summation, we measure the interference term at $2N-1$. For an integer $k$ in $[0, 2N-2]$

$$f = k f_o \tag{16}$$

Usually, the frequency sweep would not start from $f = 0$, but we ignore an offset term in equation (16) as it would only contribute to a constant phase term in equation (15) and can be omitted as a scaling factor. To comply with the Nyquist sampling condition, which is the result of the Nyquist-Shannon sampling theorem, the frequency resolution is chosen such that $f_o l_o c^{-1} = (2N-1)^{-1}$ and the spectrometer measures $PSD_{intf}(f)$ at $2N-1$ points. If $S(f)$ is known, $a(i)$ can be obtained using inverse discrete fourier transform on equation (15)

$$a(i) = \sum_{k=0}^{2N-2} \frac{PSD_{intf}(kf_o)}{S(kf_o)} \exp\left(\frac{j2\pi}{2N-1}ki\right) \tag{17}$$

$i$ for which $a(i)$ is non-zero can be used to calculate the optical distance (being $il_o$), while $a(i)$ can be used to calculate the refractive index.

*2.2. Time delay SD-OCT*

Now we place a tunable delay line in the sample arm as shown in Figure 1(b) and provide a dispersive delay that can be represented by $\zeta(f)$ in the frequency domain and has a fourier series $\theta_k$ with a period (in frequency domain) corresponding to the desired time resolution $t_o = l_o/c$. Thus equation (8) which is the field at the spectrometer becomes

$$r(t) = r_s(t) + \sum_{i=1}^{N-1} a(i) r_s(t-t_i) \otimes \sum_{k=-\infty}^{\infty} \theta_k \delta(t-t_k) \tag{18}$$

where $\otimes$ denotes convolution. The field can be written as

$$r(t) = r_s(t) + \sum_{i=1}^{N-1} a(i) \sum_{k=-\infty}^{\infty} \theta_k r_s(t-t_i-t_k) \tag{19}$$

Equation (13) is then used to calculate the autocorrelation of the interference term

$$R_{intf}(\tau) = \sum_{i=-N+1}^{N-1} a(i) \sum_{k=-\infty}^{\infty} \theta_k R_s(\tau-t_i-t_k) \tag{20}$$

The $PSD_{intf}(f)$ measured at the spectrometer is given by

$$PSD_{intf}(f) = \left[S(f) \sum_{k=-\infty}^{\infty} \theta_k \exp\left(\frac{-j2\pi kfl_o}{c}\right)\right] \sum_{i=-N+1}^{N-1} a(i) \exp\left(\frac{-j2\pi ifl_o}{c}\right) \tag{21}$$

Converting $\theta_k$ back to $\zeta(f)$ and discretizing equation (21) as before

$$PSD_{intf}(k) = \left[\zeta(k) S(k)\right] \sum_{i=-N+1}^{N-1} a(i) \exp\left(\frac{-j2\pi}{2N-1}ki\right) \tag{22}$$

Applying inverse digital fourier transform

$$u(n) = \mathbb{F}_D^{-1}\left\{(2N-1)\frac{PSD_{intf}(k)}{S(k)}\right\}(n) = h(n) \otimes a(n) \tag{23}$$

where

$$h(n) = \mathbb{F}_D^{-1}\left\{\zeta(k)\right\}(n) \tag{24}$$

Thus $h(n)$ acts like a filter on $a(n)$. Now taking the Z-transform on both sides

$$U(z) = H(z) A(z) \tag{25}$$

The capital letters $U$, $A$ and $H$ in the above equation are Z-transform, $\mathbb{Z}\{.\}$, of their corresponding small letter and $z$ is the complex variable of Z domain as shown in equation (26). Hence, the use of delay line can be interpreted as a transfer function $H(z)$ on $A(z)$ in frequency domain.

$$U(z) = \mathbb{Z}\{u(n)\} \; ; H(z) = \mathbb{Z}\{h(n)\} \; ; A(z) = \mathbb{Z}\{a(n)\} \tag{26}$$

*2.3. Cross-phase modulation SD-OCT*

Now we place the two phase modulator, one in the sample arm and the other in the reference arm as shown in Figure 1(c). The field in the sample arm is modulated by $\phi_1(t)$ while the field in the reference arm is modulated by $\phi_2(t)$. Thus, the field at the spectrometer is given by

$$r(t) = r_s(t)\exp(j\phi_2(t)) + \sum_{i=1}^{N-1} a(i) r_s(t-t_i)\exp(j\phi_1(t)) \tag{27}$$

The field in this case is no longer ergodic (not even wide sense stationary). To determine the power spectral density at the spectrometer, we will rederive Wiener-Khinchin Theorem as direct autocorrelation is unable to provide the power spectral density. Let $U_T$ be the Fourier transform $r(t)$ windowed in a region of length $T$ with center at $t = 0$.

$$U_T(f) = \int_{-T/2}^{T/2} r(t)\exp(-j2\pi f t)\, dt \tag{28}$$

Then the power spectral density is given by

$$PSD(f) = \lim_{T\to\infty} \frac{1}{T} E\left[|U_T(f)|^2\right] \tag{29}$$

where $E[.]$ is the expected value of the statistics.

$$PSD(f) = \lim_{T\to\infty} \frac{1}{T} E\left[\int_{t=-T/2}^{T/2} r(t)\exp(-2\pi f t)\, dt \int_{t'=-T/2}^{T/2} r^*(t')\exp(2\pi f t')\, dt'\right] \tag{30}$$

The interference term is given by

$$PSD_{\text{intf}}(f) = \lim_{T\to\infty} \frac{1}{T} E\left[\int_{t=-T/2}^{T/2}\int_{t'=-T/2}^{T/2} \left[\sum_{i=1}^{N-1} a(i) r_s(t-t_i) r_s^*(t')\exp(j\phi_1(t)-j\phi_2(t))\right.\right.$$
$$\left.\left. + \sum_{i=1}^{N-1} a^*(i) r_s^*(t'-t_i) r_s(t')\exp(j\phi_2(t)-j\phi_1(t))\right]\exp(-j2\pi f(t-t'))\, dt'dt\right] \tag{31}$$

$$PSD_{\text{intf}}(f) = \lim_{T\to\infty} \frac{1}{T} \int_{t=-T/2}^{T/2}\int_{t'=-T/2}^{T/2} \left[\sum_{i=1}^{N-1} a(i) E\left[r_s(t-t_i) r_s^*(t')\right]\exp(j\phi_1(t)-j\phi_2(t))\right.$$
$$\left. + \sum_{i=1}^{N-1} a^*(i) E\left[r_s^*(t'-t_i) r_s(t)\right]\exp(j\phi_2(t)-j\phi_1(t))\right]\exp(-2\pi f(t-t'))\, dt'dt \tag{32}$$

As $r_s(t)$ is an ergodic process, the expected value in the above equation is the autocorrelation function as defined by equation (2)

$$PSD_{\text{intf}}(f) = \lim_{T\to\infty} \frac{1}{T} \int_{t=-T/2}^{T/2}\int_{t'=-T/2}^{T/2} \left[\sum_{i=1}^{N-1} a(i) R_s(t-t'-t_i)\exp(j\phi_1(t)-j\phi_2(t'))\right.$$
$$\left. + \sum_{i=1}^{N-1} a^*(i) R_s(t-t'+t_i)\exp(j\phi_2(t)-j\phi_1(t'))\right]\exp(-j2\pi f(t-t'))\, dt'dt \tag{33}$$

Substituting $t = t' + \tau$ and changing limits accordingly

$$PSD_{\text{intf}}(f) = \sum_{i=1}^{N-1} a(i) \lim_{T\to\infty} \frac{1}{T} \int_{\tau=-T}^{T} R_s(\tau-t_i) \int_{t'=-T/2-\tau}^{T/2-\tau} \exp(j\phi_1(t'+\tau)-j\phi_2(t'))\exp(-j2\pi f\tau)\, dt'd\tau$$
$$+ \sum_{i=1}^{N-1} a^*(i) \lim_{T\to\infty} \frac{1}{T} \int_{\tau=-T}^{T} R_s(\tau+t_i) \int_{t'=-T/2-\tau}^{T/2-\tau} \exp(j\phi_2(t'+\tau)-j\phi_1(t'))\exp(-j2\pi f\tau)\, dt'd\tau \tag{34}$$

The second integral in the limit in both the summation terms becomes a cross-correlation function.

$$PSD_{\text{intf}}(f) = \sum_{i=1}^{N-1} a(i) \int_{\tau=-\infty}^{\infty} R_s(\tau-t_i) R_\phi(\tau)\exp(-j2\pi f\tau)\, d\tau$$
$$+ \sum_{i=1}^{N-1} a^*(i) \int_{\tau=-\infty}^{\infty} R_s(\tau+t_i) R_\phi^*(-\tau)\exp(-j2\pi f\tau)\, d\tau \tag{35}$$

where,

$$R_\phi(\tau) = \lim_{T \to \infty} \frac{1}{T} \int_{t=-T/2}^{T/2} \exp(j\phi_1(t+\tau) - j\phi_2(t)) \, dt. \tag{36}$$

The remaining integral in equation (34) resembles fourier transformation. Mapping $a(i)$ to $a(i)$ as before

$$PSD_{intf}(f) = \left[ S(f) \sum_{i=1}^{N-1} a(i) \exp(-j2\pi f t_i) \right] \otimes \Phi(f)$$
$$+ \left[ S(f) \sum_{i=-N+1}^{-1} a^*(i) \exp(-j2\pi f t_i) \right] \otimes \Phi^*(f) \tag{37}$$

where,

$$\Phi(f) = \mathbb{F}\{R_\phi(\tau)\}(f). \tag{38}$$

The second term in equation (35) is the negative part of information in length domain. If support of $\Phi^*(f)$, which acts as filter, is small compared to $N$, the second summation can be ignored in the post-processing of first summation which contains all information of $a(i)$. Thus, for the first term, the applied phase modulation results in a convolution in the frequency domain with transfer function $R_\phi(\tau)$. Discretizing the interference term and converting to Z domain gives

$$U(z) = H(z)A(z), \tag{39}$$

$$U(z) = \mathbb{Z}\{PSD_{intf}(k)\} \; ; A(z) = \mathbb{Z}\left\{ S(k) \sum_{i=1}^{N-1} a(i) \exp(-j2\pi f t_i) \right\} \; ; H(z) = \mathbb{Z}\{\Phi(k)\}. \tag{40}$$

Equation (39) represents a linear system with transfer function in time domain. The Z -transforms can be calculated by equation (40). Hence, phase modulation gives transfer function in the length domain unlike the delay line in equation (25) that results in transfer function in the frequency domain.

*2.4. Multirate Filter Bank*

Use of a tunable delay line in represent a linear system in which multirate signal processing can be used to increase the length resolution of the system as shown in Figure 1(d). Equation (25) corresponds to a transfer function block with the Z-transform in the frequency domain. As the maximum bandwidth of the spectrometer is usually limited, it may cause the resolution in the length domain (axial resolution) to be less than desired, resulting in under-sampling. Hence phase modulation can be interpreted as a transfer function [ $H(z)$ ] on the resolution limited signal [ $A(z)$ ]. Consequently, equation (25) corresponds to a single channel on the left hand side of Fig. 1(d). Likewise, multiple scans can be used to obtain information for all the channels. Next, the channels on the right-hand side are implemented on a digital computer. The depth information can be retrieved numerically by implementing the synthesis filters [ $F(z)$ ] and then combining the various channels. Similarly, for cross-arm phase modulation, equation (39) corresponds to a single channel but here the Z domain represents time domain. If the resolution of spectrometer is limited, multiple scans can be used to obtain a spectrum of desired resolution and thus providing the desired maximum depth of the OCT.

Let the spectrometer have a bandwidth that is M times smaller than required so that the axial resolution is downsampled by a factor of M from the desired $l_o$. This can be depicted by the block diagram as shown in Fig. 1(d). The block diagram resembles a single channel of the M channel filter bank. If we make the measurement M times with M different analysis filters ( $H_m$ ), the ideally sampled signal can be reconstructed using synthesis filters ( $F_m$ ). For demonstration purposes, we discuss the situation when M=2 and thus $m = [0,1]$. The perfect reconstruction (PR) of $a(n)$, which is the ideally sampled signal, is said to be achieved when $y(n) = a(n-K)$, i.e., $y(n)$ is a perfect replica of $a(n)$ and is with a shift of $K$ points. This removes both aliasing and distortion from the reconstruction. For a two-channel filter bank, the PR condition is given by

$$\begin{bmatrix} F_o(z) \\ F_1(z) \end{bmatrix} = \frac{2z^{-L}}{\Delta(z)} \begin{bmatrix} H_1(-z) \\ -H_o(-z) \end{bmatrix} \tag{41}$$

where $L=2K+1$ and $\Delta(z)$ is given by

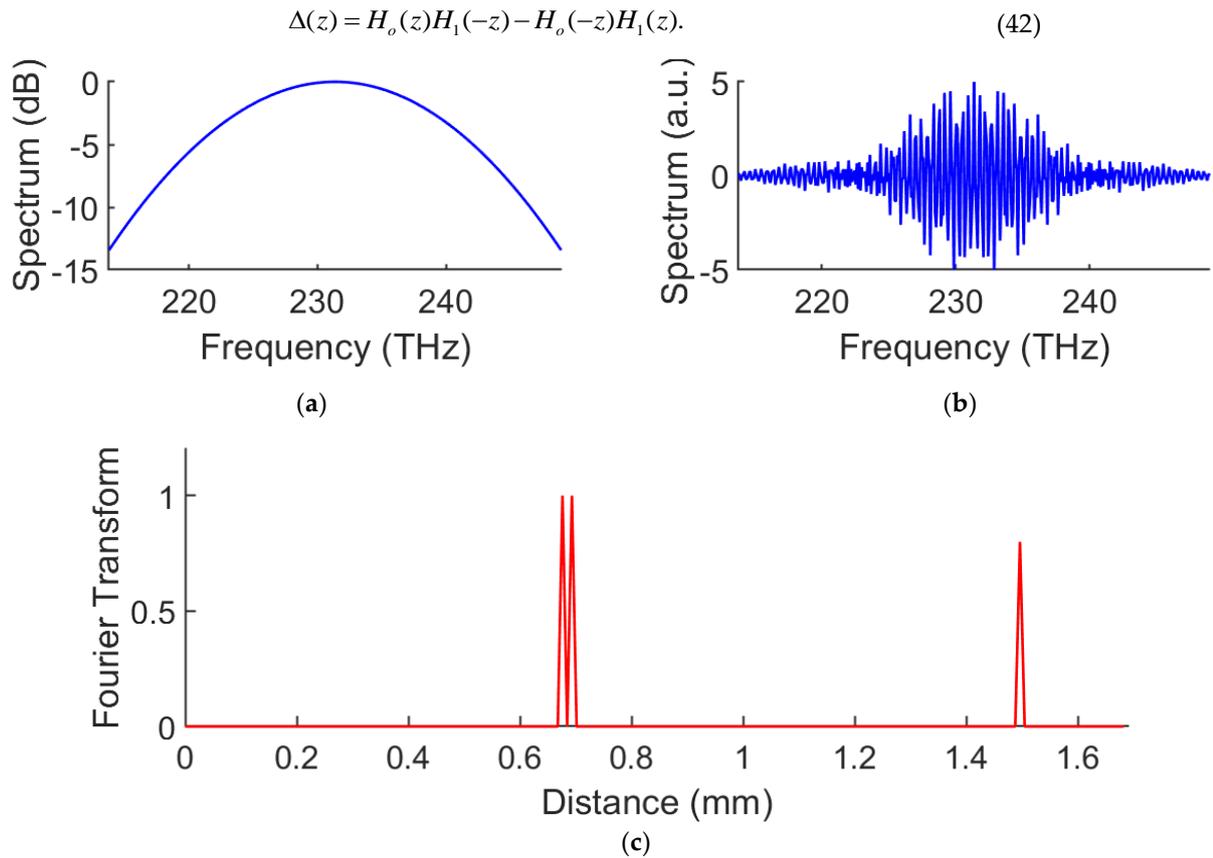

$$\Delta(z) = H_o(z)H_1(-z) - H_o(-z)H_1(z). \tag{42}$$

**Figure 2.** Demonstration of SD-OCT (**a**) Spectrum of the broadband source used for SD-OCT. (**b**) Spectrum detected by the spectrometer. (**c**) Inverse fourier transform of the detected spectrum that can be used to locate the surfaces and their thickness in the object.

Similarly, the spectrometer can have resolution M time worse than required so the frequency resolution is downsampled by M. Hence, the block diagram can be again applicable to this case but with inverted domains. Analysis filter $H_m(z)$ can be calculated using equation (26) and (40) depending on the tunable delay line/phase modulation given and synthesis filter $F_m(z)$ can be calculated using equation (41) and (42).

We stress that although equation (26) and equation (39) look the same for both the delay line and cross-arm phase modulation scenarios, it is important to know that their domain is opposite (specifically frequency and length respectively). By performing multiple scans, axial resolution can be improved in the delay line case while the maximum depth is increased in the cross arm phase modulation. Consequently, the delay is of interest when it is desired to overcome the limitation of the bandwidth of the spectrometer, while the cross-arm phase modulation is of interest when it is desired to overcome the frequency resolution limit of the system. Conceptually, these two cases are equivalent to the presence of a downsampled block in the system. Analysis filters are implemented optically using delay line/phase modulators while the synthesis filters and upsampling blocks are implemented on a digital computer. As the number of channels possible can only be integers, the resolution/maximum unambiguous range can only be improved by an integral multiple.

## 3. Simulation Results

First, to demonstrate the working of a regular SD-OCT, we simulate a single A-scan in this section. We assume a source that has a gaussian shape and is centered around 1300 nm wavelength as shown in Figure 2(a). The spectrometer used is assumed to have a bandwidth of 200 nm and 0.5 nm resolution. This corresponds to an axial optical resolution of 8.45 µm and maximum optical depth of 1.7 mm. Consider a simple object that is under investigation and is made up of three reflective surfaces. Two of them are at about optical distance of 0.69 mm with distance between them being 16.9 µm. The third surface is present at a distance of 1.5 mm. The interference spectrum that is detected at the spectrometer is given by Figure 2(b). Computing the inverse fourier transform gives the location of these surfaces in the form of peaks as shown in Figure 2(c).

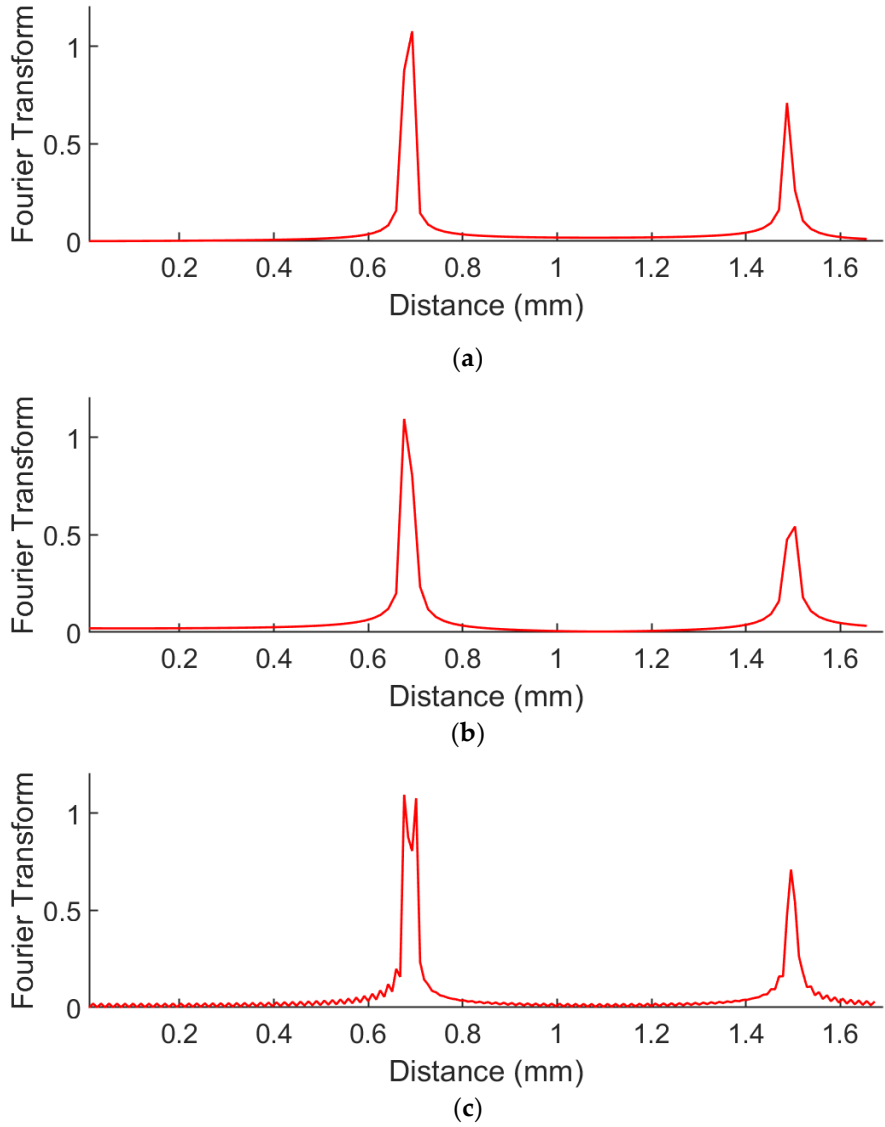

**Figure 3.** Demonstration of SD-OCT with delay line to improve axial resolution (**a**) Measured depth information from first channel. The limited bandwidth of spectrometer leads to less than desired axial resolution. Thus, the surface close together cannot be distinguished as peaks. (**b**) Measurement from the second channel that utilizes a delay line in the sample arm. (**c**) Combined graph from both the channels is able to resolve the surface close together and the position of all three surfaces are known with 2 times the accuracy compared to a single channel.

Now, we assume that the bandwidth of the spectrometer is limited to 100 nm, i.e., from 1200 nm to 1300 nm. This would mean that resolution of SD-OCT is 16.9 µm, and for the same object described above, the first two surfaces are not resolvable by the SD-OCT. This is shown in Figure 3(a) where only one peak is visible for the first two surfaces. This measurement acts like our first channel where no delay is present in the sample arm. Next, for a second channel we provide a delay given by:

$$\zeta(f) = \exp(j2\pi f t_o) \tag{43}$$

where $t_o$=28.3 fs. The length domain information of the second channel is shown by Figure 3(b). Thus, for these two channels

$$H_o(z) = 1 \,;\, H_1(z) = z^{-1}. \tag{44}$$

To fulfill the PR condition, the parameters for reconstruction can be calculated as follows

$$F_o(z) = z^{-1} \;;\; F_1(z) = 1 \;;$$
$$\Delta(z) = -2z^{-1} \;;\; K = 1 \;.$$

(45)

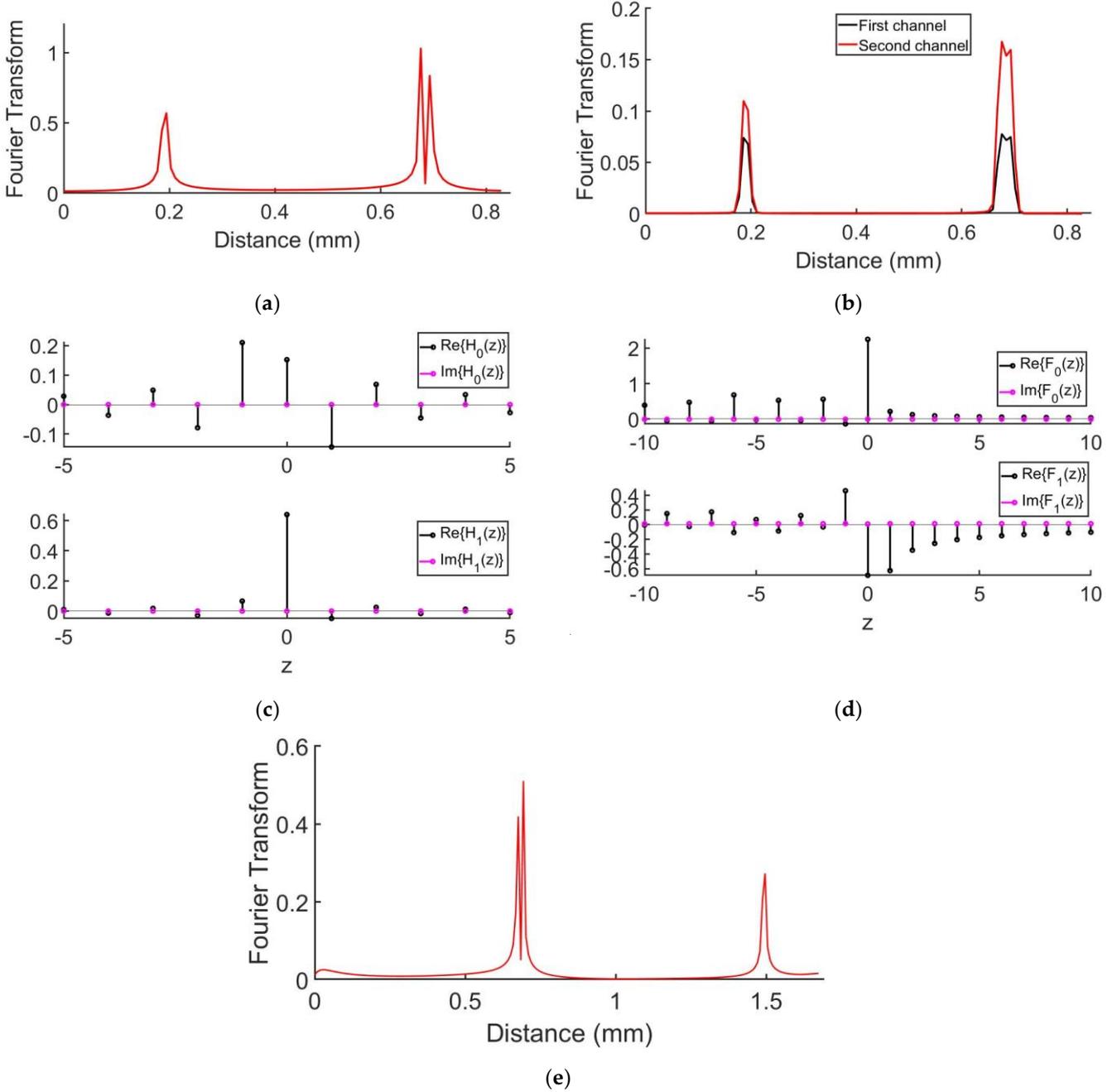

(a)

(b)

(c)

(d)

(e)

**Figure 4.** Demonstration of SD-OCT with phase modulation to improve maximum ambiguous depth (**a**) Measured depth information using a limited resolution spectrometer. The less than desired resolution results in aliasing and the location of third peak is not its true position. (**b**) The analysis filters implemented using phase modulation. (**c**) Measured depth information in the two channels after implementing phase modulation. (**d**) The synthesis filter calculated numerically using filter bank theory. (**e**) Reconstructed signal by combining the two channels after implementing synthesis filters. The maximum unambiguous range has doubled, and the measured position of the third peak is its true position.

Using both the channels and the synthesis filters, the depth information of the SD-OCT can be computed with 2 times better resolution than just a single channel. Also, the two surfaces can be resolved as the resolution has improved to distinguish their peaks. The result from combined channel is shown in Figure 3(c).

Next, we consider the case when the resolution of the spectrometer is limited to 1 nm. This results in a maximum depth of 0.85 mm and the second peak lies beyond the maximum unambiguous range. Thus, when making measurement using this limited resolution system, the aliased version of third peak appears which is not the true position of this peak. This is shown Figure 4(a) where the peak that should be at 1.5 mm appears around 0.2 mm. To increase the maximum depth of the OCT and obtain the true position of the third peak (and also the first two peaks) we use two different channels with different phase modulation. For first channel we use $\phi_{11}(t)$ and $\phi_{12}(t)$, while for the second channel we use $\phi_{21}(t)$ and $\phi_{22}(t)$. The optical phase modulation is usually generated using RF electrical signals. As arbitrary signals are difficult and cost-ineffective at high speed, we assume sinusoidal modulations. Thus, the phase modulation can be given by:

$$\phi_{pq}(t) = A_p \sin(2\pi f_{pq} t) \qquad (46)$$

where $p, q \in [1, 2]$. We choose the following values for the demonstration: $f_{11} = f_{21} = 22.1$ GHz; $f_{12} = f_{22} = 44.2$ GHz; $A_1 = 2$ rad; $A_2 = 1$ rad. The two analysis filters can be calculated by using equations (36), (39) and (40), and are shown in Figure 4(b). After the analysis filters are implemented via phase modulation, the signal is detected using spectrometer and are shown in Figure 4(c). The signal is then upsampled by a factor of 2 in frequency domain. The synthesis filters can be calculated using equation (41) and be implemented digitally on the upsampled signal. Note that to implement the synthesis filters $\Delta(z)$ should be invertible. This can be ensured by either converting $\Delta(z)$ to minimum phase filter or engineering stable synthesis filters using various modulation schemes. The coefficients of the synthesis filter are shown in Figure 4(d). After implementing the synthesis filters, the channels are combined to develop the depth information with double the maximum ambiguous length. As shown in Figure 4(e) the true position of the third surface is recovered without any aliasing.

## 4. Discussion

In summary, the multirate SD-OCT was formulated and validated with simulation. In recent literature, a number of techniques have been developed to improve the OCT either using superior hardware [38, 39] or complex post-processing [40, 41]. The novelty of our multirate SD-OCT is not only that it combines the effect of both additional hardware and post processing but also is compatible with some of the other existing technique as use of filters and multiple channels is universally applicable to a linear system. Also, some of the techniques previously shown in literature such as use of multiple broadband sources [30] to improve axial resolution is a special case of multirate SD-OCT where each broadband source can be considered as separate channel multiplexed in wavelength rather in time as in our case.

The resolution and bandwidth of the SD-OCT system are often limited by the spectrometer. Grating spectrometers are widely used in SD-OCT [42] and its performance is determined by various physical parameters such as grating length, number of gratings, material used, wavelengths, etc. [43]. Often these parameters are interdependent and are restricted by technology and feasibility. This, in turn, makes the resolution and bandwidth of the spectrometer interdependent and limited. By using multirate SD-OCT, not only the cost of spectrometer is reduced for one of bandwidth and resolution, but it also gives the opportunity sacrifice one for the other as the sacrificed parameter can be improved by this technique. The price to pay for this technique is the modulator/delay line cost and the extra time required to carry out multiple scans. As the object in SD-OCT are usually stationary, extra scans would not pose a challenge.

**Author Contributions:** Conceptualization, P.G. and A.G.; methodology, P.G.; software, P.G.; validation, P.G., A.G. and Y.F.; formal analysis, P.G.; investigation, P.G.; resources, Y.F.; data curation, P.G.; writing—original draft preparation, P.G.; writing—review and editing, A.G. and Y.F.; visualization, P.G.; supervision, A.G. and Y.F.; project administration, Y.F.; funding acquisition, Y.F. All authors have read and agreed to the published version of the manuscript.

**Funding:** This work was partially supported by the National Science Foundation (NSF) grant NSF ECCS-2023730, the San Diego Nanotechnology Infrastructure (SDNI) supported by the NSF National Nanotechnology Coordinated Infrastructure (grant ECCS-2025752), and the ASML/Cymer Corporation.

**Data Availability Statement:** The data presented in this study are available on request from the corresponding author.

**Acknowledgments:** We acknowledge Alexander Franzen for providing ComponentLibrary

**Conflicts of Interest:** The authors declare no conflict of interest.